\documentclass[twocolumn,showpacs,preprintnumbers,amsmath,amssymb]{revtex4}

\usepackage{epsf}
\usepackage{graphicx}
\usepackage{latexsym}
\usepackage{amssymb}
\usepackage{amsfonts}
\usepackage{soul}
\usepackage{color}
\begin{document}
\bibliographystyle{apsrev}

\title{Hybridization and interference effects for localized superconducting states in strong magnetic field}
\author{A.~Yu.~Aladyshkin, A.~S.~Mel'nikov, I.~M.~Nefedov, D.~A.~Savinov,
M.~A.~Silaev, I.~A.~Shereshevskii}
\affiliation{Institute for
Physics of Microstructures RAS, 603950, Nizhny Novgorod, GSP-105,
Russia}
\date{\today}

\begin{abstract}
Within the Ginzburg -- Landau model we study the critical field
and temperature enhancement for crossing superconducting channels
formed either along the sample edges or domain walls in thin-film
magnetically coupled superconducting -- ferromagnetic bilayers.
The corresponding Cooper pair wave function can be viewed as a
hybridization of two order parameter (OP) modes propagating along
the boundaries and/or domain walls. Different momenta of
hybridized OP modes result in the formation of vortex chains
outgoing from the crossing point of these channels. Near this
crossing point the wave functions of the modes merge giving rise
to the increase in the critical temperature for a localized
superconducting state. The origin of this critical temperature
enhancement caused by the wave function squeezing is illustrated
for a limiting case of approaching parallel boundaries and/or
domain walls. Using both the variational method and numerical
simulations we have studied the critical temperature dependence
and OP structure vs the applied magnetic field and the angle
between the crossing channels.
\end{abstract}



\maketitle

\section{Introduction}

Recent experimental and theoretical studies of
ferromagnet/superconductor (F/S) heterostructures have revealed a
rich physics of magnetic and transport properties of these systems
(see, e.g., Refs.~\cite{Buzdin,Aladyshkin} for review). A
considerable amount of attention in these studies has been devoted
to the effect of ferromagnetic domain structure on the critical
temperature of superconductivity nucleation (see, e.g.,
Refs.~\cite{Kopaev,Buzdin_2,Buzdin_3,Aladyshkin_2,Rusanov,Yang}
and references therein). This effect originates from both the
exchange and electromagnetic (orbital) mechanisms of interaction
between superconducting and magnetic orderings. The
electromagnetic mechanism appears to be a dominant one for the
experimental works which investigate F/S bilayers designed to
suppress the proximity effect (see, e.g.,
Refs.~\cite{Yang,Werner}). The non-uniform magnetic stray field of
domain walls can result in the formation of a localized Cooper
pair wave function at temperatures exceeding the superconducting
critical temperature in the bulk. Such localized order parameter
(OP) distributions and corresponding phase diagrams have been
studied in Refs.~\cite{Buzdin_2,Aladyshkin_2,Buzdin_3} for a
generic case of a step-like profile of a stray magnetic field in a
superconducting thin film. At a fixed temperature $T$ for a
certain critical amplitude $B_0$ [$H_{c2}(T)<B_0\leq H_{c3}(T)$]
of a step -- like magnetic stray field profile the structure of
localized OP wave function coincides with the one which formes in
the vicinity of superconductor/vacuum or superconductor/insulator
flat interface in a parallel field equal to $B_0$, where
$H_{c2}(T)$ and $H_{c3}(T)=1.695H_{c2}(T)$ are the critical fields
of bulk superconductivity and surface superconductivity,
respectively~\cite{Saint-James}.

Certainly, in a restricted sample geometry the localized OP wave
function should appear not only near the domain walls but also at
the film edges. Thus, for temperature values $T$ between
$T_{c2}(B_0)$ and $T_{c3}(B_0)>T_{c2}(B_0)$, defined by the
conditions $H_{c2}(T_{c2})=B_0$ and $H_{c3}(T_{c3})=B_0$,
respectively, there appears a set of localized superconducting
modes propagating along the sample edges and/or domain walls and
decaying in the transverse direction. Hereafter these localized
modes will be called superconducting channels. One can pose a
natural question whether the superconducting critical temperature
$T_c$ could be increased due to the overlapping of the OP modes
localized near the various types of superconducting channels.
Indeed, such increase in the $T_c$ value is well known to occur
for two parallel superconductor/vacuum interfaces (i.e. for a
superconducting film) placed in a homogeneous magnetic field $H$
parallel to the interfaces (see, e.g.,
Refs.~\cite{{Saint-James,SaintJames-1969}}). The critical
temperature in this system monotonously increases with the film
thickness decreasing and saturates at the $T_{c0}=T_{c}(H=0)$
value in the limit of small thicknesses. Taking non-parallel
boundaries we get a superconducting wedge placed in a magnetic
field. The squeezing of the superconducting wave function at a
small wedge angle is known to cause a strong increase in the
critical temperature $T_c$ (see
Refs.~\cite{Mel'nikov,Brosense,Gelder,Houghton,Klimin,Schweigert,Fomin,Bonnaillie,Bonnaillie_2,Bonnaillie_3}).

It is the goal of our work to analyze the effect of such wave
function squeezing on the critical temperature enhancement for
various types of crossing superconducting channels localized near
the sample edges and/or domain walls. Within the linearized
Ginzburg -- Landau model we carried out the calculations of the
critical temperature of localized superconductivity for a simple
hybrid system consisting of a thin superconducting film placed in
a non-uniform stray magnetic field of straight domain boundaries
in a ferromagnetic substrate underneath. An origin of the critical
temperature enhancement can be clearly demonstrated for a generic
example of parallel boundaries. Generalizing the textbook solution
for the critical field of a thin superconducting film
\cite{{Saint-James,SaintJames-1969}} we consider two model
problems: (i) parallel domain boundary and sample boundary
separated by the distance $D$; (ii) two parallel domain boundaries
at the separation $D$. In both cases the dependence of critical
temperature $T_c$ vs the distance $D$ reveals a maximum at the
value of the order of the magnetic length $L_B=\sqrt{\hbar
c/|e|B_0}$, where $\hbar$ is a Planck constant, $e=-|e|$ is an
electron charge. At this distance the wave functions of the OP
modes localized near neighboring domain walls or near the domain
wall and the sample edge merge and form a single superconducting
channel. Turning now to the case of non--parallel domain walls
(which cross at the angle $\varphi_0\ll\pi/2$) or a domain wall
crossing a sample boundary at a certain small angle $\varphi_0$
one can expect that the most favorable conditions for
superconductivity nucleation realize at the distance $\sim
L_B\varphi_0$ from the crossing point. As a result, for rather
small angles $\varphi_0\ll\pi/2$ the center of the energetically
favorable OP distribution is shifted from the crossing point. This
shift is accompanied by the striking phenomenon: the critical
temperature $T_c$ of the localized superconductivity monotonously
increases when the tilting angle $\varphi_0$ tends to zero. Such
behavior makes the dependence $T_c(\varphi_0)$ nonanalytic at
$\varphi_0=0$ where the critical temperature exhibits a jump to
the value $T_{c3}$ ($T_{c2}$) for the case of the domain wall
crossing the sample edge (for two crossing domain walls).

In our work we suggest a simple variational procedure which allows
us to get an approximate solution of a linearized Ginzburg --
Landau equation describing the hybridization of the localized
superconducting states for an arbitrary $\varphi_0$ angle. We find
out that different momenta of hybridized OP modes are responsible
for the formation of vortex chains outgoing from the crossing
point of the channels. The effect of these chains on the OP trial
function appears to be extremely important for rather large
$\varphi_0$ angles close to $\pi/2$. The important effect of such
vortex chain can be illustrated for the simplest example of the
superconductivity nucleation at the wedge corner. Our calculations
show that only taking into account the vortex chain one can get a
proper crossover to the $H_{c3}$ field with the increase in the
wedge angle up to the flat one. The solution of a generic problem
describing the superconductivity nucleation in a wedge allows us
to find appropriate trial wave functions for superconducting OP
nucleating near the domain walls intersecting the film edge or
near the crossing point of domain boundaries. The change in the
$\varphi_0$ angle is found to be accompanied by the change in the
orientation of vortex chains and the intervortex distance. Our
analytical findings based on the variational procedure are in a
good agreement with direct numerical simulations.

Applying an external magnetic field perpendicular to the film
plane one can observe the increase in the critical temperature of
the domain wall superconductivity due to the partial magnetic
field compensation inside the domains. Using both variational and
numerical approaches we have composed the phase diagram of the F/S
bilayer in the plane temperature $T$ -- external magnetic field
$H$ and discuss the transitions between different superconducting
states by varying temperature $T$, external magnetic field $H$ and
magnetic stray field amplitude $B_0$.

The paper is organized as follows. In Sec.~II we mainly focus on
the variational analysis of the superconductivity nucleation and
interference of localized superconducting states. In Sec.~III we
present the results of direct numerical simulations which support
our analytical findings.  Finally, the results are summarized in
Sec.~IV.

\section{Variational approach}
We start our study of the hybridization and interference effects
for interacting superconducting channels forming in a thin-film
magnetically coupled F/S bilayer with the GL variational procedure
focusing on the analysis of appropriate trial wave functions. Let
us analyze the problem of the OP nucleation in a thin
superconducting film placed in the non-uniform magnetic field
${\bf B}({\bf r})=H\mathbf z_0 +{\bf b}({\bf r})$ induced by the
external sources and the magnetic domain walls in a ferromagnetic
substrate, respectively. Note that we will restrict ourselves to
the case of a step-like distribution of the magnetic stray fields
of the domain walls and neglect the effect of the magnetic field
components parallel to the film plane. In particular, for a single
domain wall we take ${\bf b}({\bf r})=\mathbf z_0B_0{\rm
sign}(\widetilde{x})$, where the $\widetilde{x}$ axis is directed
perpendicular to the corresponding domain wall. Thus, we neglect
the deviations from this step-like field model caused by a finite
thickness of a superconducting film and decay of the magnetic
stray field at large distances from the domain walls (see
discussion in
Refs.~\onlinecite{Aladyshkin_2,Aladyshkin_3,Aladyshkin_4}).

\subsection{Linearized Ginzburg -- Landau model}
The superconducting critical temperature $T_c$ can be routinely
determined from the linearized GL equation
    \begin{equation}
    \label{Eq:GL-vec}
    \left[-i\nabla-\frac{2e}{\hbar c}{\bf
    A}({\bf r})\right]^2\Psi({\bf r})= \frac{1}{\xi^2(T)}\Psi({\bf r})
    \end{equation}
as the highest possible value: $T_c = {\rm max}~T$, corresponding
to the lowest "energy level" $E\propto \xi^{-2}(T)$ of the
eigenvalue problem~(\ref{Eq:GL-vec}). Here $\Psi({\bf r})$ is the
OP distribution, ${\bf A}({\bf r})$ is the vector potential
corresponding to the total magnetic field ${\bf B}({\bf r})$, $m$
is the electron mass, $\xi(T)=\xi_0/\sqrt{1-T/T_{c0}}$ is the
superconducting coherence length. Alternatively, $T_c$ is known to
be determined from the variational problem
    \begin{eqnarray}
    \frac{1}{\xi^2(T)}=\,\frac{\int |(-i\hbar\nabla-
    2e{\bf A(\mathbf r)}/c)\Psi(\mathbf r)|^2d^2\mathbf r}
    {\hbar^{2}\int |\Psi(\mathbf r)|^2d^2\mathbf r} \ ,
    \label{functional GL}
    \end{eqnarray}
and the integration is performed over the superconducting volume.
The wave function $\Psi(\mathbf r)$ satisfies the boundary
condition
    \begin{eqnarray}
    \left.\big[-i\hbar\frac{\partial}{\partial\mathbf n}-
    \frac{2e}{c}\mathbf A_n(\mathbf r)\big]\Psi(\mathbf r)\right|_{\rm\Gamma} =
    0 \ ,
    \label{BoundCond-3}
    \end{eqnarray}
where $\mathbf n$ is a unit vector normal to the sample boundary
$\rm\Gamma$.

The complex-valued wave function $\Psi(\bf r)$ can be written in
the form $\Psi(\mathbf r)=f(\mathbf r)e^{i\Theta(\mathbf r)}$,
where $f(\mathbf r)$ and $\Theta(\mathbf r)$ are the absolute
value and the phase of $\Psi(\mathbf r)$, respectively. Thus, we
can rewrite Eq.~(\ref{functional GL}) in the form
    \begin{eqnarray}
    \frac{1}{\xi^2(T)}=\frac{\int \Big\{4m^2v_s^2(\mathbf r)f^2(\mathbf r)+\hbar^2[\nabla f(\mathbf r)]^2\Big\}d^2\mathbf r}
    {\hbar^2\int f^2(\mathbf r)d^2\mathbf r} \ ,
    \label{functional GL2}
    \end{eqnarray}
where $\mathbf v_s(\mathbf r)=[\hbar\nabla\Theta(\mathbf
r)-2e\mathbf A(\mathbf r)/c]/2m$ is a superfluid velocity.

\subsection{Localized superconducting modes forming in the presence of parallel boundaries}
We begin with the consideration of the superconductivity
nucleation for parallel sample edges and/or domain walls in a
thin-film F/S bilayer. Let us choose the $x$ ($y$) axis to be
perpendicular (parallel) to these boundaries and take the gauge
${\bf A}=A_y(x)\,{\bf y}_0$. The Schr\"odinger--like equation
(\ref{Eq:GL-vec}) does not depend on $y$ coordinate and one can
generally find the solution in the form
$\Psi(x,y)=f_k(x)\exp(-iky)$, where the function $f_k(x)$ should
be determined from the following eigenvalue problem:
    \begin{eqnarray}
    \label{Eq:GL-x}
    - \frac{d^2 f_{k}(x)}{dx^2} +  U(x) f_{k}(x) =
    \frac{1}{\xi^2(T)}f_{k}(x) \ , \quad
    \\
    \nonumber
    U(x)=\left[\frac{2\pi}{\Phi_0}A_y(x)-k\right]^2.
    \end{eqnarray}

    \begin{figure}[htb!]
    \includegraphics[height=0.35\textwidth]{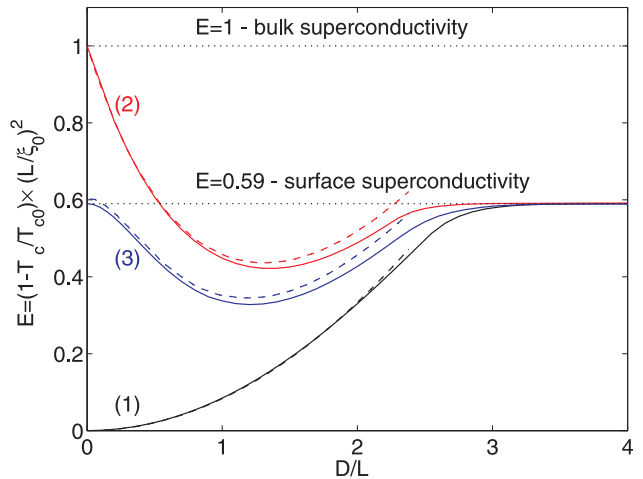}
    \caption{(Color online) The shift of the energy
    $E=(1-T_c/T_{c0})\times (L/\xi_0)^2$ vs the distance $D$,
    obtained numerically (solid lines) and using the trial function
    approach (dashed lines). The curves (1), (2), (3) correspond to three model problems: (i) two
    superconductor/vacuum interfaces , (ii) two domain walls, (iii) domain wall and superconductor/vacuum interface,
    respectively. Here $L=L_H$ for the first problem, $L=L_B$ for the last two problems.}
    \label{Fig:LGL-Spectra}
    \end{figure}
We calculate the dependence of the shift of the critical
temperature $(1-T_c/T_{c0})$ on the distance $D$ using both
numerical solution of the eigenvalue problem (\ref{Eq:GL-x}) and
variational approach.

{\it Two superconductor/vacuum interfaces.} For two parallel
superconductor/vacuum interfaces positioned at $x=\pm D/2$ and
forming a superconducting slab of the finite thickness $D$ placed
in a homogeneous parallel magnetic field $B_z=H$ we should apply
the boundary condition $df_k(x)/dx|_{x=\pm D/2}=0$. The dependence
of $E=(1-T_c/T_{c0})\,(L_H/\xi_0)^2$ on $D$ (where
$L_H=\sqrt{\hbar c/|eH|}$ is a characteristic length scale in a
magnetic field), originally calculated by Saint-James and de
Gennes~\cite{{Saint-James,SaintJames-1969}} is shown in
Fig.~\ref{Fig:LGL-Spectra} by the solid line~(1). The critical
temperature $T_c$ tends to $T_{c0}$ at $D\ll L_H$ and, thus, the
magnetic field has a negligible effect on the OP nucleation for
rather small $D$ values. In this limit a simple approximation for
the dependence $E(D)$ can be found provided we choose $A_y(x)=Hx$
and assume the OP wave function to be uniform across the slab. The
minimum in the dependence $E(k)$ in this case corresponds to
$k=0$. Such approximation while being rigorously justified only
for $D\ll L_H$ appears to describe the monotonous decrease in
$T_c$ with the increasing $D$ distance even for $D\sim L_H$ (see
the dashed line~(1) in Fig.~\ref{Fig:LGL-Spectra}). In the limit
$D\gg L_H$, there are two symmetrical minima in the dependence
$E(k)$ positioned at nonzero $k$ values. These two solutions
correspond to the same critical temperature $T_{c3}$ of surface
superconductivity: $T_c\simeq T_{c3}=T_{c0}\,(1-0.59
|H|/H_{c2}^{(0)})$, where $H_{c2}^{(0)}=\Phi_0/2\pi\xi_0^2$ is the
upper critical field at $T=0$, and $\Phi_0=\pi\hbar c/|e|$ is the
magnetic flux quantum. The corresponding OP wave functions
describe the localized superconducting modes running along the
boundaries $x=-D/2$ and $x=D/2$.

{\it Two domain walls.} We proceed our consideration with the case
of two parallel domain walls separated by the distance $D$: $B_z=
-B_0$ for $-D/2\le x \le D/2$ and $B_z=B_0$ elsewhere (here we
assume $H=0$). In the limit $D=0$ the critical temperature $T_c$
equals to the bulk critical temperature $T_{c2}=
T_{c0}\,(1-B_0/H_{c2}^{(0)})$ in the uniform magnetic field $B_0$.
In the opposite limiting case $D\gg L_B=\sqrt{\hbar c/|e|B_0}$ the
OP wave function localized at the discontinuities of the magnetic
field component $B_z$ saturates the critical temperature $T_{c3}$.
What is less intuitively clear is that the transition from
$T_{c2}$ to $T_{c3}$ with  the increasing distance $D$ occurs via
the $T_c$ enhancement. Such non-monotonous $T_c$ behavior can be
captured with a good accuracy by minimizing the energy
functional~(\ref{functional GL2}) for the Gaussian-like trial
function: $f_k(x)=\exp(-x^2/a^2)$, where $a$ is a variational
parameter. The choice $A_y(-x)=-A_y(x)$ automatically yields
$k=0$. Then it is clear that the rise in $T_c$ is caused by the
increase in the width of the effective potential well
$U(x)=A^2_y(x)$ for increasing $D$ and by the lowering of the
ground energy level in a wider potential well. The comparison of
the results of the numerical solution of Eq.~(\ref{Eq:GL-x})
[solid line~(2)] with the trial function approach [dashed
line~(2)] is presented in Fig.~\ref{Fig:LGL-Spectra}.

{\it Domain wall and superconductor/vacuum interface.} Finally, we
analyze the case of a single domain wall parallel to the
superconducting film edge and positioned at a distance $D$ from
this edge. We also assume $H=0$ so that the only characteristic
length scale is $L_B$. Similarly to the previous case the $T_c$
value changes non-monotonuously as a function of $D$ and in both
limits ($D\ll L_B$ and $D\gg L_B$) the critical temperature tends
to the critical temperature of surface superconductivity:
$T_c\rightarrow T_{c3}$. At $D\rightarrow \infty$ there are two
independent superconducting OP nuclei located near the surface
($x=-D/2$) and at the domain wall ($x=D/2$) both characterized by
the same critical temperature $T_{c3}$. Analogously to the
previous case of two parallel domain walls the non-monotonous
behavior of $T_c$ is caused by the increasing width of the
potential well $U(x)=\left[2\pi A_y(x)/\Phi_0-k\right]^2$ in the
Eq.~(\ref{Eq:GL-x}). The problem can be apparently mapped on the
one in the infinite superconducting slab in the magnetic field of
three parallel domain walls placed at $x=\pm D/2$ and $x=-3D/2$.
The resulting magnetic field distribution is an odd function for
reflection respective to the plane $x=-D/2$. For such magnetic
field configuration the ground state solution should possess the
reflection symmetry: $f_k(x-D/2)=f_k(-x-D/2)$. Therefore for
$D\leq L_B$ we can choose the trial function in the form
$f_k(x)=\exp[-(x-D/2)^2/a^2]$, where $a$ is a variational
parameter. Unlike the previous case the vector potential now is an
even function $A_y(x-D/2)=A_y(-x-D/2)$ which means that the ground
state solution corresponds to $k\neq 0$. Minimizing the energy
functional over the parameters $a$ and $k$, we obtain the shift of
the critical temperature shown by dashed line~(3) in
Fig.~\ref{Fig:LGL-Spectra}. This plot again demonstrates an
excellent agreement with the numerical result [solid line~(3)] for
rather small $D$ values.

Thus the solution of these model problems allows us to make an
important observation about the possibility to get the critical
temperature enhancement for a pair of approaching boundaries or
domain walls.

\subsection{Hybridization of localized superconducting modes
propagating along the channels crossing at small angles}
Considering the problem of superconductivity nucleation for
crossing boundaries or domain walls it is natural to start from
the simplest case of small crossing angles when the distance $D$
between the crossing boundaries changes adiabatically.

{\it Two superconductor/vacuum interfaces.} The solution for two
crossing superconductor/vacuum interfaces which form a
superconducting wedge with small corner angle $\chi\ll\pi/2$  can
be found in
Refs.~\cite{Mel'nikov,Brosense,Gelder,Houghton,Klimin,Schweigert,Fomin,Bonnaillie,Bonnaillie_2,Bonnaillie_3}
Let us introduce a cylindrical coordinate system $(r,\varphi,z)$.
The monotonous increase in $T_c$ with the decreasing distance $D$
discussed above allows to assume that the maximum of the OP wave
function should be positioned at the wedge vertex ($r=0$).
Substituting the simplest isotropic wave function $\Psi({\bf
r})=\exp(-r^2|eH|\chi/2\sqrt{3}\hbar c)$ in the
functional~(\ref{functional GL}) and carrying out the variational
procedure one can find the following asymptotical expression for
the critical magnetic field $H_{c3}^{\rm w}\simeq
\sqrt{3}H_{c2}/\chi$ suppressing the localized superconductivity.
    \begin{figure}[h]
    \begin{center}
    \includegraphics[width=8.5cm]{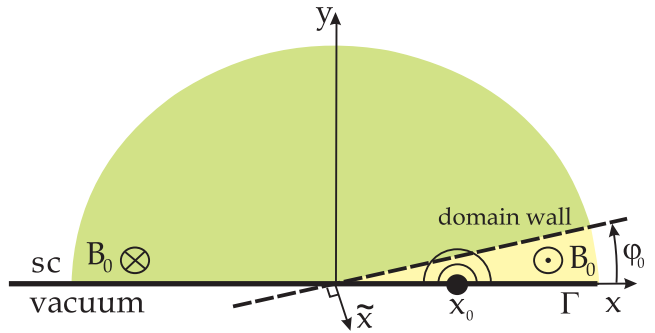}
    \end{center}
    \caption{(Color online) Schematic view of a F/S
    bilayer: a semi-infinite thin superconducting film and
    a straight domain wall (dashed line) oriented at a certain angle
    $\varphi_0\ll\pi/2$ with respect to the film edge $\Gamma$. $B_0$ is a
    stray field amplitude inside the domain.}
    \label{Small_angle}
    \end{figure}

{\it Domain wall crossing the superconductor/vacuum interface.} We
proceed now with the case of a single domain wall oriented at a
rather small angle $\varphi_0$ with respect to the superconducting
film edge $\Gamma$ (see Fig.~\ref{Small_angle}). We restrict
ourselves by a particular case of zero external field $H=0$. Thus,
we consider the variational problem (\ref{functional GL}),
(\ref{BoundCond-3}) for a semi-infinite superconducting thin film
in a magnetic stray field $b_z(\mathbf r)=B_0{\rm
sign}(\widetilde{x})$. We choose the trial function $f(\mathbf r)$
in the form:
    \begin{eqnarray}
    \label{Psi_small_angle}
    f(\mathbf r)=e^{-\delta y^2}e^{-\beta (x-x_0)^2},
    \end{eqnarray}
where $\delta$, $\beta$ and $x_0$ are the variational parameters.
Here we allow the center of a superconducting nucleus to be
shifted from the crossing point of the domain wall and the sample
edge $\Gamma$ only along the $x$ axis ($x_0\neq 0$). The shift of
the nucleus center is a direct consequence of the existence of the
maximum in the dependence of the critical temperature $T_c$ vs the
separation $D$ between the boundaries. According to the above
treatment this maximum corresponds to the $D$ value of the order
of the magnetic length $L_B$. Thus, for a small angle $\varphi_0$
the peak in the OP wave function should appear at the distance
$\sim L_B\varphi_0$ from the crossing point.

Accounting the magnetic stray field $b_z(\mathbf r)$ of the domain
wall we take the superfluid velocity $\mathbf v_s(\mathbf r)$ in
each magnetic domain to be equal to the vector potential $\mathbf
A(\mathbf r)$ chosen in the radial gauge:
    \begin{eqnarray}
    \label{VP}
        \nonumber
        \mathbf v_s(\mathbf r)=B_0r(\varphi-\vartheta)\times |e|/(mc)\,\mathbf r_0 \ ,
        \\
        \nonumber
        \quad 0\leq\varphi<\varphi_0 \ , \\
        \nonumber
        \mathbf v_s(\mathbf r)=B_0r(-\varphi+2\varphi_0-\vartheta)\times |e|/(mc)\,\mathbf r_0 \ ,
        \\
        \quad \varphi_0<\varphi\leq\pi \ ,
    \end{eqnarray}
where $\vartheta$ is the variational parameter ($0\leq
\vartheta\leq\varphi_0$).
    \begin{figure}[htb!]
    \includegraphics[width=8cm]{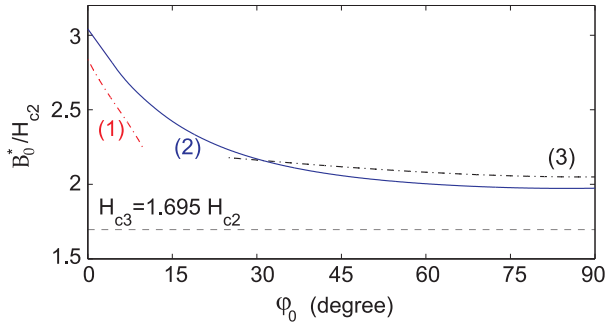}
    \caption{(Color online) The critical amplitude $B_0^{*}$
    of the domain stray field of localized superconductivity nucleation in a thin-film
    semi-infinite F/S bilayer as a function of the titling angle $\varphi_0$. Dash-dot lines (1) and (3) are obtained
    from the variational analysis carried out both for small and large corner angles, respectively.
    The dependence $B_0^{*}(\varphi_0)$ found from numerical simulations is shown by the solid line~(2).
    } \label{Angular_dependence}
    \end{figure}
    \begin{figure}[htb!]
    \begin{center}
    \includegraphics[height=3.8cm]{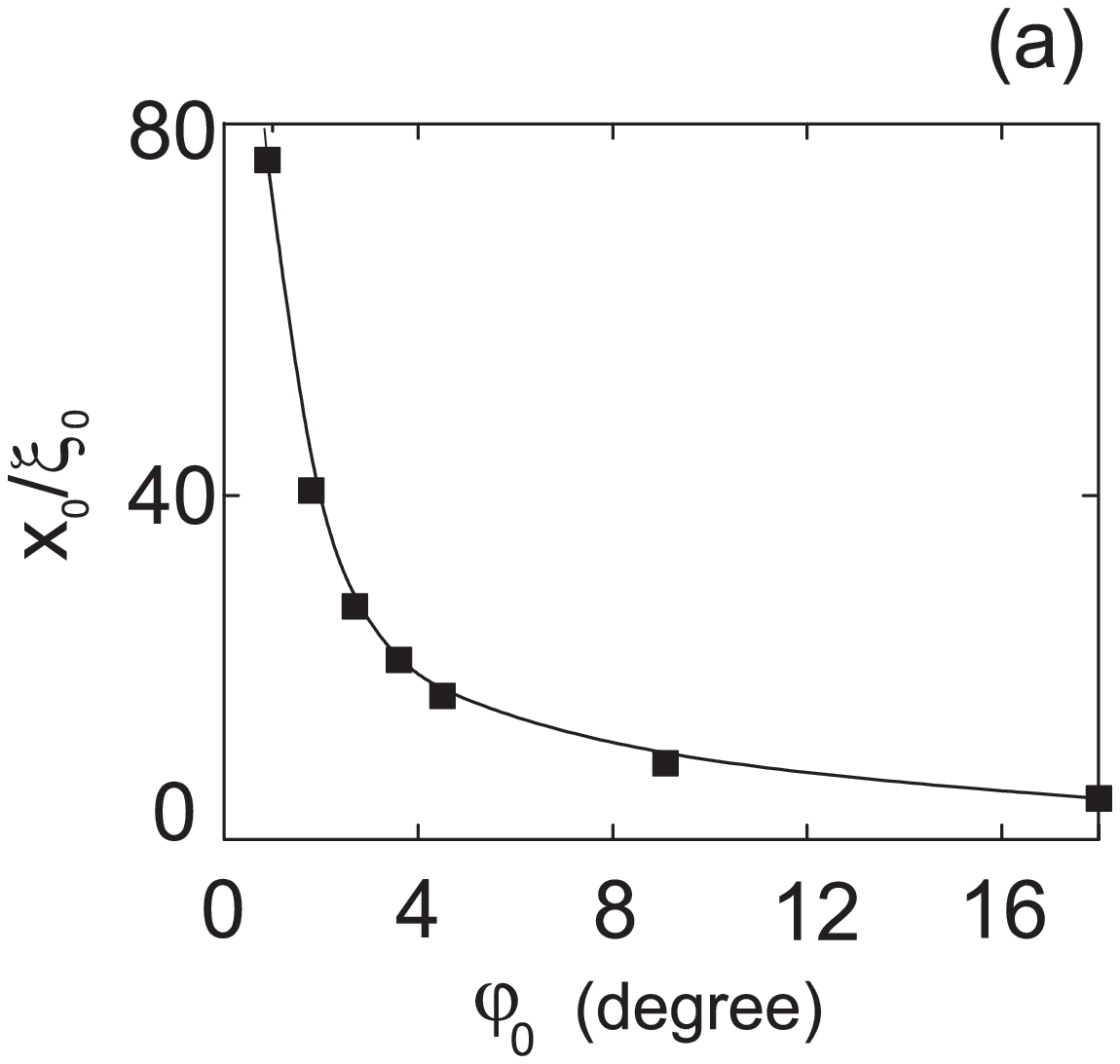}
    \includegraphics[height=3.8cm]{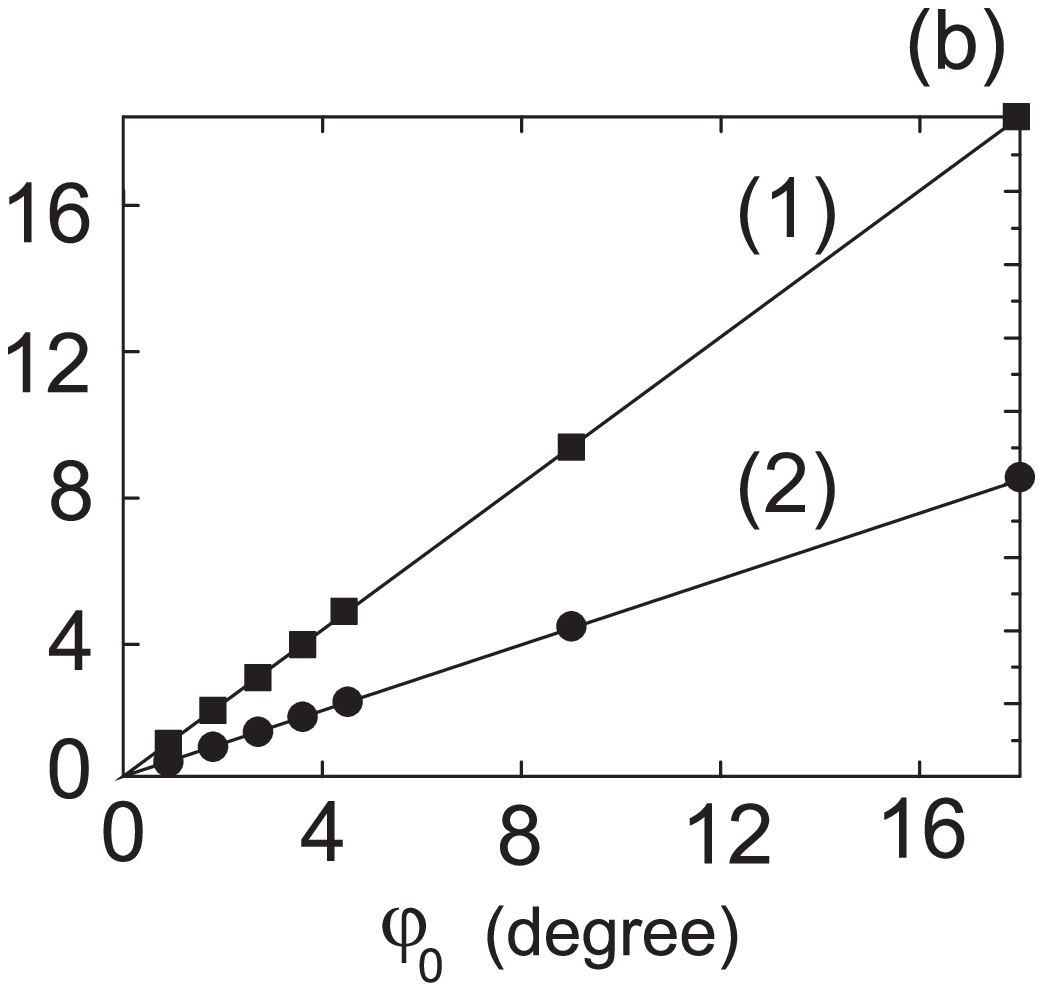}
    \end{center}
    \caption{(a) Variational parameter $x_0$ vs $\varphi_0$ obtained
    using the trial function approach (squares). The solid line corresponds to the
    analytical dependence $x_0=0.4\pi/\varphi_0$. (b) The dependences of $\beta\times 180^o$ (squares) and
    $\vartheta$ (circles) vs $\varphi_0$ obtained using the trial function approach.
    The solid lines (1) and (2) correspond to the analytical dependences $\beta\times 180^o=\varphi_0$
    and $\vartheta=\varphi_0 /2.2$, respectively.}
    \label{variational parameters}
    \end{figure}
    \begin{figure}[htb!]
    \begin{center}
    \includegraphics[width=8.5cm]{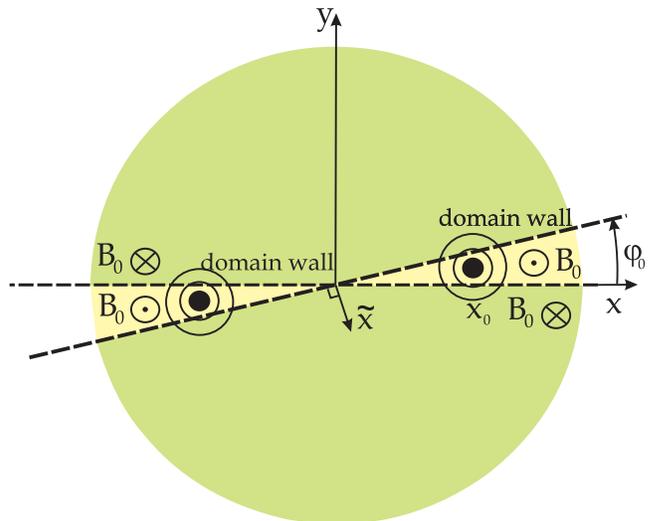}
    \end{center}
    \caption{(Color online) Schematic view of a F/S
    bilayer: a thin superconducting film and
    two domain walls (dashed lines) crossing at a certain angle
    $\varphi_0\ll\pi/2$. $B_0$ is a stray field amplitude inside the domain.}
    \label{Small_angle_domain_walls}
    \end{figure}
Substituting the expressions~(\ref{Psi_small_angle}) and
(\ref{VP}) into the Eq.~(\ref{functional GL2}) and carrying out
the minimization over the  parameters $\delta$, $\beta$, $x_0$,
and $\vartheta$, we derive the dependence of the critical field
amplitude $B_0^{*}$ vs the $\varphi_0$ angle for the limit of
small angles $\varphi_0\ll 1$. Typical plot of this dependence is
shown in Fig.~\ref{Angular_dependence} by a dash-dot line~(1). For
$\varphi_0\ll 1$ this plot demonstrates a good agreement with our
numerical simulations [see the solid line~(2) in
Fig.~\ref{Angular_dependence}] carried out below. It is
interesting to note that the position $x_0$ of the OP maximum
diverges inversely proportional to $\varphi_0$ as $\varphi_0\to 0$
in accordance with our expectation [see Fig.~\ref{variational
parameters}(a)]. The parameter $\delta$ is found to be independent
of $\varphi_0$: $\delta =0.16\xi_0$. The dependences of the
variational parameters $\beta$ and $\vartheta$ vs the $\varphi_0$
angle are presented in Fig.~\ref{variational parameters}(b).

One can see that even for $\varphi_0\ll 1$ the critical field
$B_0^{*}$ do differs from the $H_{c3}$ value corresponding to
conventional surface superconductivity in a homogeneous field:
$B_0^{*}\rightarrow 2.8 H_{c2}$ at $\varphi_0\rightarrow 0$.  At
$\varphi_0=0$ the dependence $B_0^{*}(\varphi_0)$ exhibits a
step-like jump to the $H_{c3}$ field. This jump should be, of
course, smeared for a finite size sample when $x_0$ becomes
comparable to the sample size.

It is important to note that the threshold value $B_{0}^{*}$
allows to determine the critical temperature at $H=0$:
$T_c=T_{c0}\,(1-B_0/B^{*}_0).$

{\it Two domain walls.} The consideration in previous subsections
can be easily generalized for the case of two domain walls which
cross at rather small angle $\varphi_0\ll\pi/2$ (see
Fig.~\ref{Small_angle_domain_walls}). Due to the symmetry of the
magnetic field distribution we can consider only the half-space:
$0\le \varphi\le \pi$. We take the superfluid velocity $\mathbf
v_s(\mathbf r)$ in the form (\ref{VP}) and choose the trial
function $f(\mathbf r)$ like in the following
    \begin{eqnarray}
    f(\mathbf r)=e^{-\delta (y-y_0)^2}e^{-\beta
    (x-x_0)^2} \ .
    \label{Psi_small_angle_2}
    \end{eqnarray}
Here $\delta$, $\beta$, $x_0$ and $y_0 $ are the variational
parameters. It is important to note that the center of a
superconducting nucleus is shifted from the crossing point of two
domain walls along both the $x$ and $y$ axes. Substituting the
expressions~(\ref{VP}) and (\ref{Psi_small_angle_2}) into the
Eq.~(\ref{functional GL2}) and carrying out the minimization over
the parameters $\delta$, $\beta$, $x_0$, $y_0$, and $\vartheta$,
we derive the dependence $B_0^{*}(\varphi_0)$ for the limit of
small angles $\varphi_0\ll 1$. This critical field dependence
appears to be very close to the one shown in
Fig.~\ref{Angular_dependence} by the dash-dot line~(1): the
relative deviation is of the order of several percents. We
emphasize that the nucleus center $(x_0, y_0)$ shifts along the
bisectrix of the $\varphi_0$ angle.

\subsection{Interference of localized superconducting modes propagating along the channels crossing at large angles}
We proceed now with the consideration of the localized
superconductivity nucleation for the superconducting channels
crossing at rather large angles.

{\it Two superconductor/vacuum interfaces.} We start with an
exemplary problem of superconductivity nucleation in a wedge with
the corner angle $\chi\leq\pi$ placed in a uniform magnetic field
$H$. The superconducting wave function localized near the wedge
vertex can be considered as an overlapping of the superconducting
modes propagating along the wedge sides. These modes are
characterized by the wave vectors parallel to the different sides
and, thus, the interference of these localized waves should result
in a formation of vortices at the bisectrix of the wedge angle.
The superconductivity in a wedge appears for $H>H_{c3}$ and, thus,
the OP decays with the increasing distance from the wedge vertex.
    \begin{figure}[htb!]
    \begin{center}
    \includegraphics[width=8.5cm]{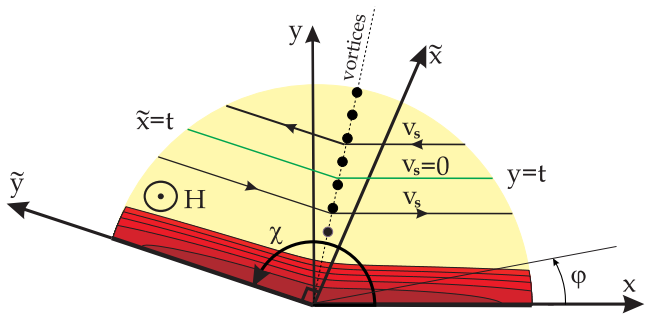}
    \end{center}
    \caption{(Color online) Cross section of a superconducting wedge
    with an arbitrary corner angle $\chi$ in a uniform magnetic field
    $H$. The solid circles show the vortex chain outgoing from the wedge vertex along its
    bisectrix. The lines of a superfluid velocity $\mathbf v_s$ and
    the contour lines of $f(\mathbf r)$ are presented. Dashed lines
    ($\widetilde{x}=t$ and $y=t$) correspond to $\mathbf v_s=0$.}
    \label{Wedge}
    \end{figure}
We would like to note that the interference effect and the
resulting formation of a vortex chain have been disregarded in
previous
works~\cite{Mel'nikov,Brosense,Gelder,Houghton,Fomin,Jadallah}
considering different types of trial functions for the OP in a
wedge. As a consequence, all these variational calculations
provided poor agreement with the numerical results (see, e.g.,
Ref.~\cite{Schweigert}) for rather large wedge angles up to $\pi$.
We will demonstrate that only accounting the vortex chain outgoing
from the wedge vertex one can obtain a proper crossover to the
$H_{c3}$ field at $\chi\rightarrow\pi$.

Considering a superconducting wedge (see Fig.~\ref{Wedge}) and
introducing the dimensionless variables in the
functional~(\ref{functional GL2}) we come to the following
expression
    \begin{eqnarray}
    \frac{1}{h}=
    \frac{\int \{\mathbf v'^2_s(\mathbf r')f^2(\mathbf r')+[\nabla f(\mathbf r')]^2\}d^2\mathbf r'}
    {\int f^2(\mathbf r')d^2\mathbf r'} \ ,
    \label{functional GLH}
    \end{eqnarray}
where $h^{-1}=L_H^2/\xi_0^2\times\Big(1-T/T_{c0}\Big)$,
$d^2\mathbf r'=L_H^{-2}d^2\mathbf r$, $\mathbf v'_s(\mathbf
r')=\nabla\Theta(\mathbf r')+a(\mathbf r')$ is a dimensionless
superfluid velocity and $a(\mathbf r')=|H|^{-1}\mathbf A(\mathbf
r')$ is a dimensionless vector potential.

Similarly to Ref.~\cite{Mel'nikov} we choose the absolute value
$f(\mathbf r')$ of Cooper pair wave function $\Psi(\mathbf r')$ in
the form
    \begin{eqnarray}
    f(\mathbf r')=\exp\{-\alpha r'^2[1+\gamma\sin^2(\pi\phi/\chi)]\} \ ,
    \label{Psi modulus}
    \end{eqnarray}
where $\alpha$ and $\gamma$ are the variational parameters,
$r'=L_{H}^{-1}r$, $\phi=\varphi-\chi/2$. To describe the vortex
chain positioned along the bisectrix we introduce a cut along the
line $\phi =0$ where the superfluid velocity experiences a
discontinuity. The tangential jump in the $v_s$ value corresponds
to a continuously distributed vorticity along this cut. For this
purpose we divide the wedge into the angular domains
$-\chi/2\leq\phi<0$ and $0\leq\phi\leq\chi/2$ and take the
dimensionless superfluid velocity $v'_s(\mathbf r')$ in each
domain equal to the vector potential $a(\mathbf r')$ chosen
parallel to the wedge sides at $\phi = \mp\chi/2$,
correspondingly. Both gauges of the vector potential $\mathbf
a(\mathbf r')$ correspond to a uniform field $\mathbf{H}$:
\begin{eqnarray}\label{dimensionless superfluid velocity}
        \nonumber
           &&\mathbf v'_s(\mathbf r')=(t-y')\,\mathbf x_0 \ ,\quad -\chi/2\leq\phi<0 \ , \\
           &&\mathbf v'_s(\mathbf r')=(\widetilde{x}-t)\,\widetilde{\mathbf y}_0 \ ,\quad 0\leq\phi\leq\chi/2
           \ ,
\end{eqnarray}
where $t$ is a variational parameter,
$(\widetilde{x},\widetilde{y})$ is a new reference system rotated
at the angle $\chi-\pi/2$ with respect to the original system
$(x,y)$ in the counter-clockwise direction (see Fig.~\ref{Wedge}).
Such choice of the superfluid velocity provides a correct
asymptotical behavior of the OP modes propagating along different
wedge sides at large distances from the vertex. Of course, a
single-valued wave function should vanish at the cut and, thus,
the choice of superfluid velocity in the form
Eq.~(\ref{dimensionless superfluid velocity}) is not adequate for
description of the wave function behavior close to the wedge
vertex ($r\lesssim\xi$) where the absolute value of the OP  is
essentially nonzero for all angles $\phi$. As a consequence, we
can use this method only for wave functions strongly elongated
along the wedge sides when the region close to the vertex provides
a small contribution to the functional (\ref{functional GLH}). We
will see below that this condition appears to break down for small
angles $\chi$ when the wave function is almost isotropic for all
distances $r$.

\begin{figure}[h]
    \begin{center}
    \includegraphics[width=7.cm]{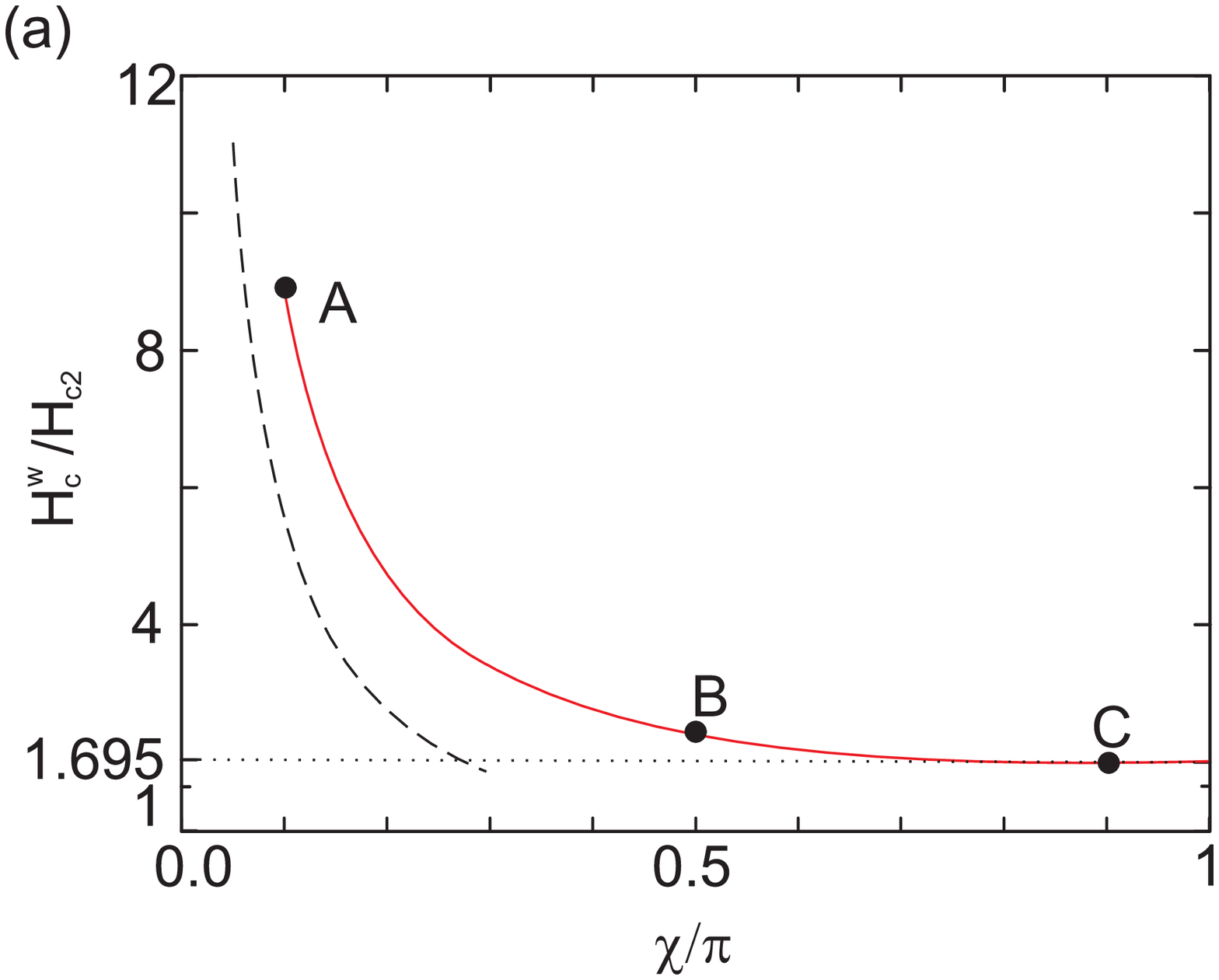}
    \includegraphics[width=7.cm]{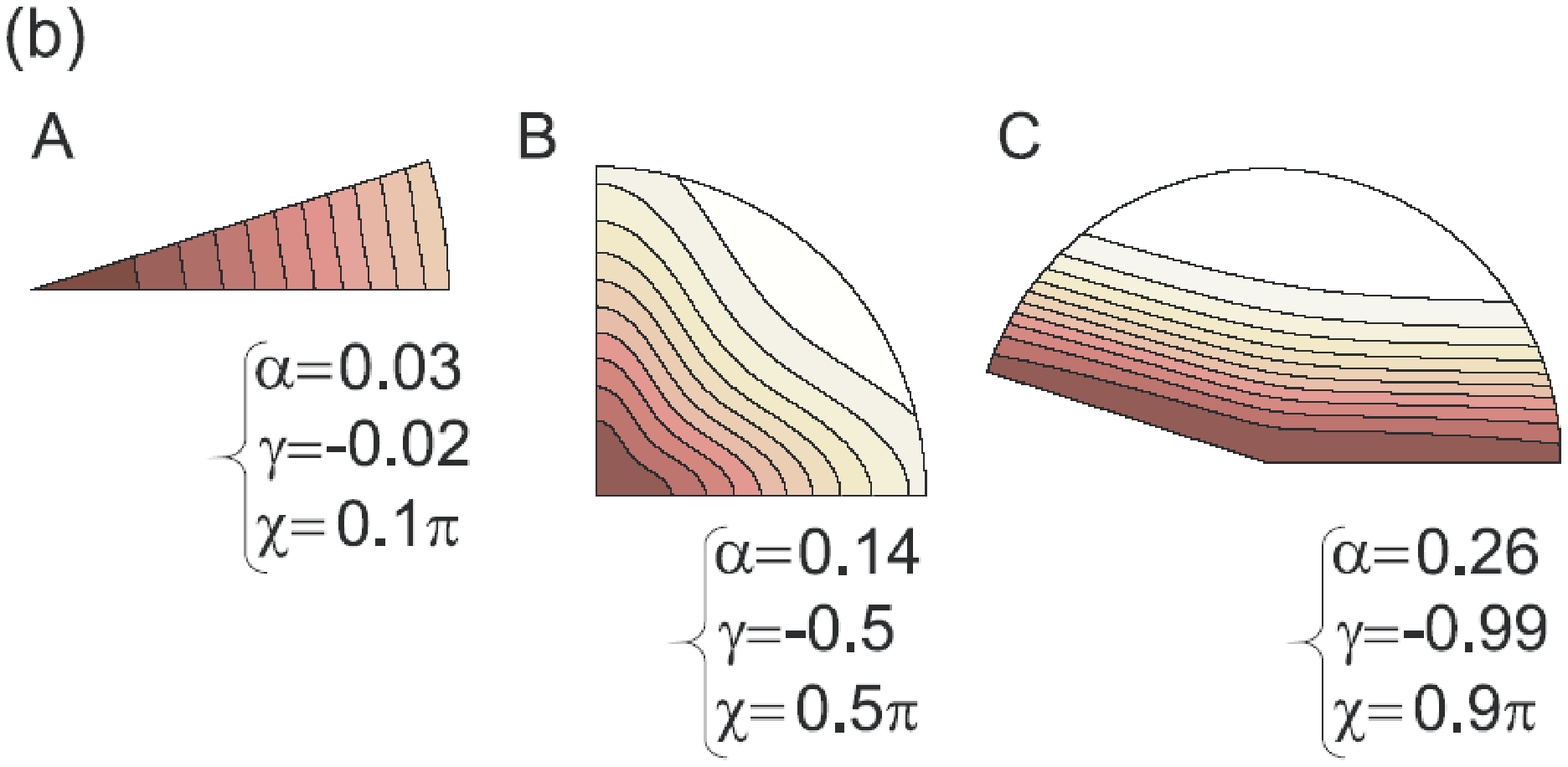}
    \end{center}
    \caption{(Color online) (a) Typical plots of the nucleation field $H_{c3}^{\rm
    w}$ in a wedge vs the corner angle $\chi$. The solid line shows
    the results of our variational calculations while the dashed line
    corresponds to asymptotic behavior of the nucleation field
    $H_{c3}^{\rm w}$ at small corner angles $\chi\ll 1$ according to
    Refs.~\cite{Mel'nikov,Brosense,Gelder,Houghton,Klimin,Schweigert,Fomin,Bonnaillie,Bonnaillie_2,Bonnaillie_3}.
    (b) The contour plots of $f(\mathbf r')$ function inside the
    areas $r'\leq 5$ for different $\chi$ angles: $0.01\pi$ (A),
    $0.5\pi$ (B) and $0.9\pi$ (C).} \label{nucleation field in a wedge}
    \end{figure}

Rewriting the expressions~(\ref{dimensionless superfluid
velocity}) in terms of $r'$ and $\phi$ and substituting them into
the Eq.~(\ref{functional GLH}) one obtains the function
$h^{-1}=h^{-1}(\alpha,\gamma,t,\chi)$. Carrying out the
minimization over the variational parameters $\alpha$, $\gamma$
and $t$ we find the critical field $H_{c3}^{\rm
w}(T)=H_{c2}(T)\times[\min_{\alpha,\gamma,t} (1/h)]^{-1}$ of
superconductivity nucleation for different wedge angles $\chi$.
Typical plot of the dependence $H_{c3}^{\rm w}(\chi)$ is presented
in Fig.~\ref{nucleation field in a wedge} by the solid line.

One can see that for $\chi=\pi$ we obtain the result found
previously by Saint-James and de Gennes~\cite{Saint-James}:
$H_{c3}^{\rm w}(\pi)\simeq 1.695H_{c2}$. For a particular case
$\chi=\pi/2$ we find $H_{c3}^{\rm w}(\pi/2)\simeq 2H_{c2}$. This
value appears to be in a good agreement with the numerical
calculations carried out in Ref.~\cite{Schweigert} (see also the
comparison with numerical simulations below). The variational
parameters corresponding to this case should be taken as follows:
$\alpha(\pi/2)=0.14,\, \gamma(\pi/2)=-0.5,\, t(\pi/2)=0.6$.

Analyzing the typical contour plots of the $f(\mathbf r')$
function in the insets of Fig.~\ref{nucleation field in a wedge}
for different $\chi$ one can see that the angular anisotropy of
the OP vanishes ($\gamma\rightarrow 0$) in the limit
$\chi\rightarrow 0$. For such small angles $\chi$ we find out:
$H_{c3}^{\rm w}/H_{c2}\simeq 1.56\sqrt{3}\, \chi^{-1}$. This
asymptotical behavior deviates from the correct dependence
$H_{c3}^{\rm w}/H_{c2}=\sqrt{3}\,\chi^{-1}$ found previously in
Refs.~\cite{Mel'nikov,Brosense,Gelder,Houghton,Klimin,Schweigert,Fomin,Bonnaillie,Bonnaillie_2,Bonnaillie_3}.
This deviation for small angles is a natural consequence of the
wave function isotropy as it is discussed above. With the increase
in the wedge angle $\chi$ the anisotropy parameter $|\gamma|$
grows and the wave function becomes elongated along the wedge
sides which restores the validity of our approach. The vortex --
free trial functions considered in
Refs.~\cite{Mel'nikov,Brosense,Gelder,Houghton,Fomin,Jadallah} can
no more provide a correct behavior of the upper critical field
$H_{c3}^{\rm w}$.

\begin{figure}[h]
    \begin{center}
    \includegraphics[width=8.5cm]{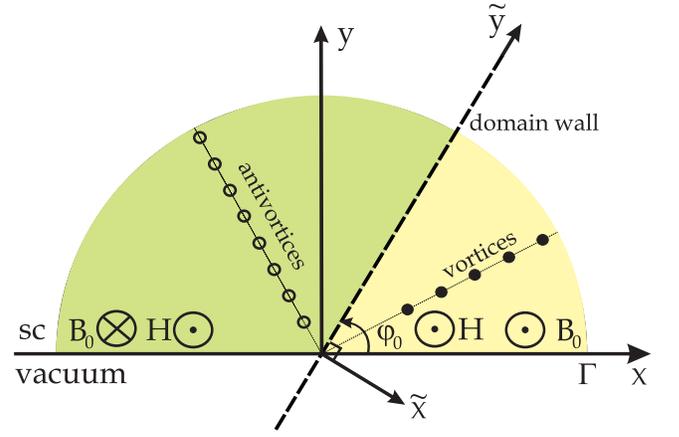}
    \end{center}
    \caption{(Color online) Schematic view of a F/S
    bilayer: a semi-infinite thin superconducting film and
    a straight domain wall (dashed line) oriented at a certain angle
    $\varphi_0$ with respect to the film edge $\Gamma$. $B_0$ is a
    stray field amplitude inside the domain,
    $H$ is an external magnetic field, the $\widetilde{x}$
    axis is chosen to be perpendicular to the domain wall.
    The vortex chain is shown by the solid circles while
    the opened circles indicate the antivortices.}
    \label{System_variational_analysis}
    \end{figure}

{\it Domain wall crossing the superconductor/vacuum interface.}
The variational procedure described in the latter subsection can
be easily generalized for other model systems containing crossing
superconducting channels. We proceed now with the consideration of
the localized superconductivity nucleation in a thin-film
semi-infinite F/S bilayer with a domain wall crossing the sample
edge $\Gamma$ at rather large angle $\varphi_0\leq\pi/2$ (see
Fig.~\ref{System_variational_analysis}).
    \begin{figure*}[htb!]
    \begin{center}
    \includegraphics[width=8.5cm]{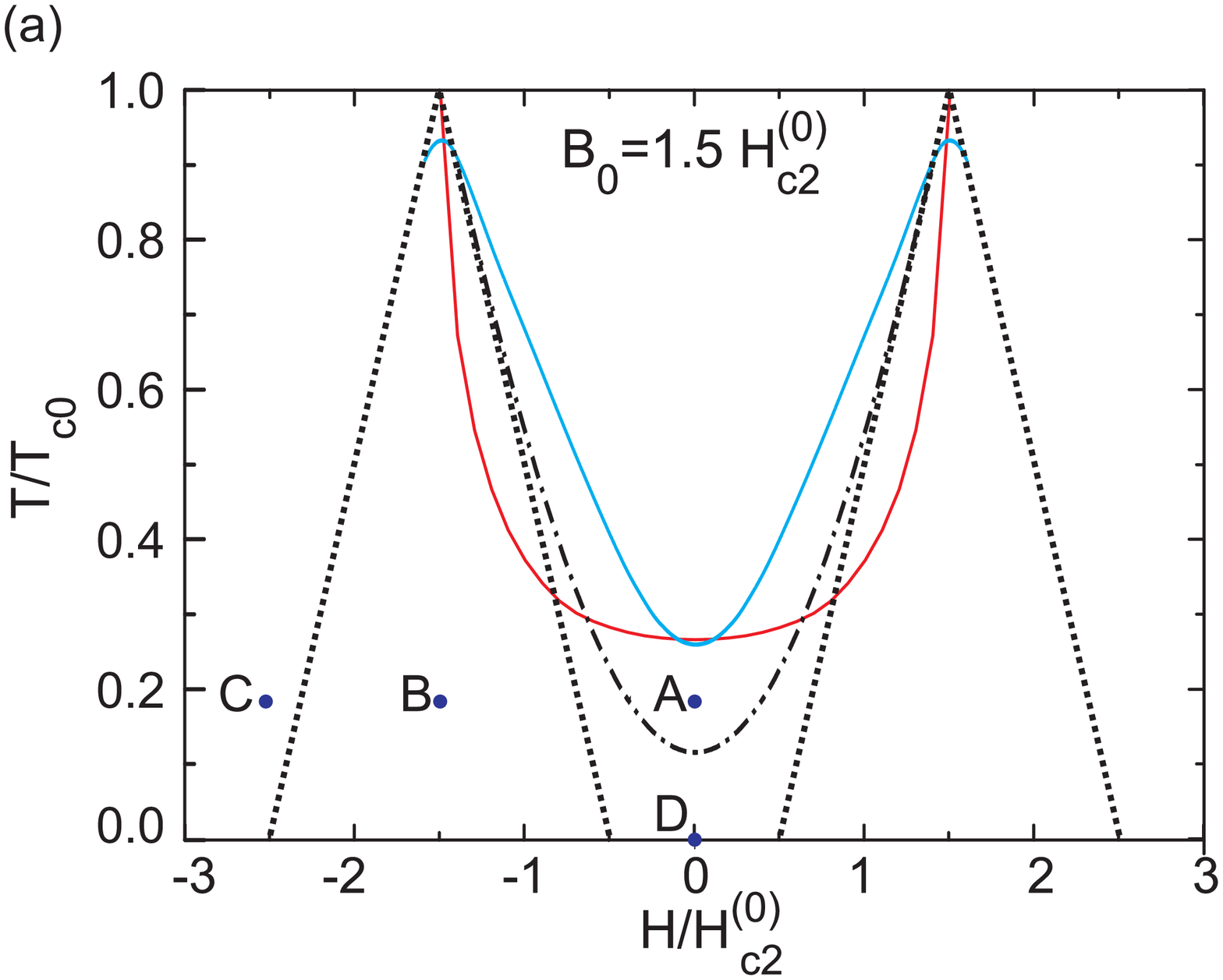}
    \epsfxsize=170mm \epsfbox{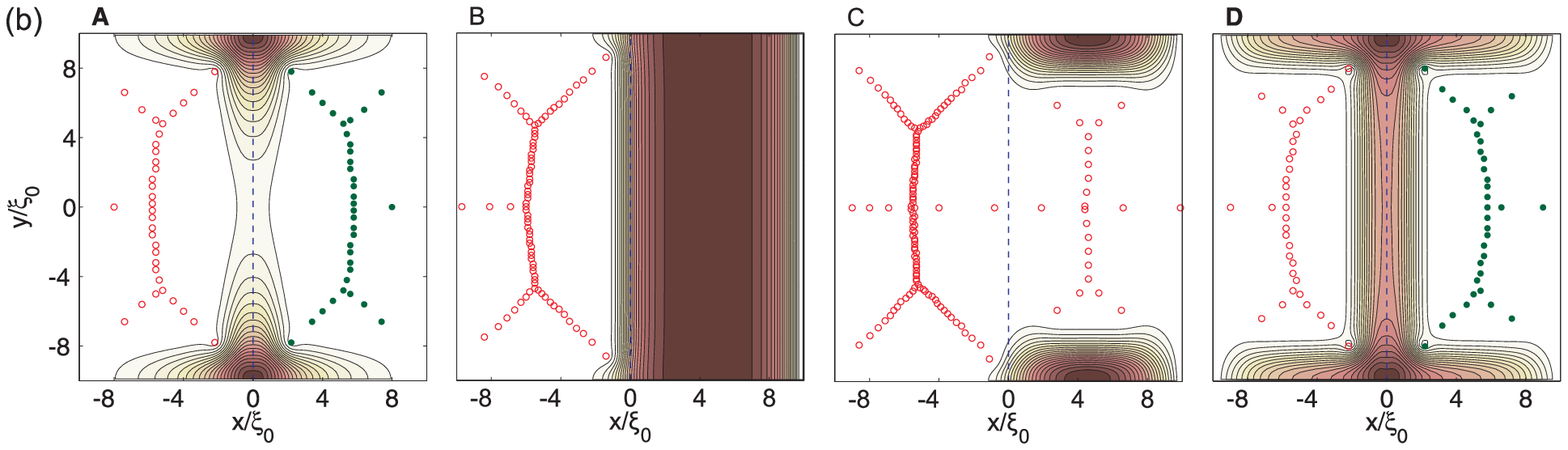}
    \end{center}
    \caption{(Color online) (a) Typical plots of the critical
    temperature $T_{c3}^{*}(H)$ for localized superconductivity in a thin-film
    semi-infinite F/S bilayer with a
    domain wall oriented perpendicular to the sample edge $\Gamma$ ($\varphi_0=\pi/2$,
    see Fig.~\ref{System_variational_analysis}). The
    solid line shows the results of our variational calculations
    while the dashed line corresponds to our numerical
    simulations carried out for a rectangular superconducting film
    with lateral dimensions $20\xi_0\times 20\xi_0$.
    The dotted lines correspond to the dependence of the shifted upper critical field of a bulk
    superconductor vs temperature: $T=T_{c0}\left\{1-\big||H|-B_0\big|/H_{c2}^{(0)}\right\}$,
    the dash-dot line shows the phase-transition
    line for a domain-wall superconductivity derived, e.g., in
    Refs.~\cite{Buzdin_3,Aladyshkin_2} Here we take $B_0=1.5H_{c2}^{(0)}$.
    (b) Typical plots of superconducting OP distributions presented for different points
    on the $H$ -- $T$ plane: A, B, C and D,
    where the dashed line indicates the domain wall. The vortex chains are shown by the solid
    circles while the antivortices are presented by open circles.}
    \label{Phase Diagram}
    \end{figure*}
To generalize the considered approach one needs to introduce two
vortex chains outgoing from the crossing point of a domain wall
and the sample edge $\Gamma$. It is naturally to expect these
vortex chains to be oriented along the bisectrices of two
wedge-shaped regions located to the left ant to the right of the
domain wall. We follow the procedure described in the previous
subsection and replace the vortex chains by the cuts for
superfluid velocity, i.e., by the lines with a continuously
distributed vorticity. The  absolute value of the OP wave function
and the dimensionless superfluid velocity are taken in the forms:
    \begin{eqnarray} \label{Psi modulus EA_DWS}
        \nonumber
        f_1(\mathbf r)=\exp\{-\alpha_1
        r^2[1+\gamma_1\cos^2(\pi\varphi/\varphi_0)]\} \ ,
        \\
        \nonumber
        0\leq\varphi<\varphi_0 \ , \\
        \nonumber
        f_2(\mathbf r)= \\
        \nonumber
        \exp\Big\{-\alpha_2 r^2\Big[1+\gamma_2\cos^2\Big(\pi(\varphi-\varphi_0)/(\pi-\varphi_0)\Big)\Big]\Big\} \ ,
        \\
        \varphi_0\leq\varphi\leq\pi \ ,
    \end{eqnarray}
    \begin{eqnarray} \label{Superfluid velocity EA_DWS}
        \nonumber
        \mathbf v_s(\mathbf r)=(H+B_0)
        (\eta-y)\times |e|/(mc)\,\mathbf x_0 \ ,
        \\
        \nonumber
        0\leq\varphi<\varphi_0/2 \ , \\
        \nonumber
        \mathbf v_s(\mathbf r)=(H+B_0)(\widetilde{x}-\eta)\times |e|/(mc)\,\widetilde{\mathbf
        y}_0 \ ,
        \\
        \nonumber
        \varphi_0/2\leq\varphi<\varphi_0 \ ,
        \\
        \nonumber
        \mathbf v_s(\mathbf r)=(H-B_0)(\widetilde{x}+\eta)\times |e|/(mc)\,\widetilde{\mathbf
        y}_0 \ ,
        \\
        \nonumber
        \varphi_0\leq\varphi<\pi/2+\varphi_0/2 \ ,
        \\
        \nonumber
        \mathbf v_s(\mathbf r)=(H-B_0)(\eta-y)\times |e|/(mc)\,\mathbf x_0 \ ,
        \\
        \pi/2+\varphi_0/2\leq\varphi\leq\pi \ ,
    \end{eqnarray}
where $\alpha_1$, $\gamma_1$, $\alpha_2$, $\gamma_2$, and $\eta$
are the variational parameters.  The continuity condition for the
wave function $\Psi(\mathbf r)$ at $\varphi=\varphi_0$ gives the
relation
    \begin{eqnarray}
    \alpha_1(1+\gamma_1)=\alpha_2(1+\gamma_2) \ .
    \label{continuity condition}
    \end{eqnarray}

For simplicity we start with a particular case of the domain wall
perpendicular to the film edge $\Gamma$ ($\varphi_0=\pi/2$).
Substituting the expressions~(\ref{Psi modulus EA_DWS}) and
(\ref{Superfluid velocity EA_DWS}) into the
functional~(\ref{functional GL2}) and carrying out the
minimization over the variational parameters $\alpha_1$,
$\gamma_1$, $\alpha_2$, $\gamma_2$, and $\eta$ at a fixed
amplitude $B_0$ of the magnetic stray field  we find the critical
temperature $T^{*}_{c3}$ vs the applied field $H$. Typical plot of
the phase-transition line $T^{*}_{c3}(H)$ is presented in
Fig.~\ref{Phase Diagram} by a solid line for
$B_0=1.5H^{(0)}_{c2}$. One can see that applying an external
magnetic field $H$ we obtain an increase in the critical
temperature $T^{*}_{c3}$ of localized superconductivity due to the
partial magnetic field compensation inside the domains.

As a next step we restrict ourselves to the case $H=0$ and analyze
the dependence of the critical amplitude $B_0^{*}$ of the domain
stray field corresponding to the superconductivity nucleation vs
the $\varphi_0$ angle. Substituting the expressions~(\ref{Psi
modulus EA_DWS}) and (\ref{Superfluid velocity EA_DWS}) into the
Eq.~(\ref{functional GL2}) we carry out the minimization procedure
over the variational parameters $\alpha_1$, $\gamma_1$,
$\alpha_2$, $\gamma_2$, and $\eta$. Typical plot of the dependence
$B_0^{*}(\varphi_0)$ shown in Fig.~\ref{Angular_dependence} by a
dash-dot line~(3) is in a good agreement with the solid line~(2)
derived within our numerical simulations discussed below.

It is important to note that with the decrease in the angle
$\varphi_0$ we clearly observe an increase in the field $B_0^{*}$.
This increase obviously occurs due to a partial shrinking of the
Cooper pair wave function in analogy to the case of a
superconducting wedge considered above. For small angles
$\varphi_0\rightarrow 0$ we find: $B_0^{*}/H_{c2}\rightarrow 2.2$.
In the case $\varphi_0=\pi/2$ for an appropriate gauge choice the
problem of a half -- infinite domain wall can be exactly mapped to
the wedge problem for $\chi=\pi/2$. Indeed, for a domain wall
perpendicular to the edge $\Gamma$ the GL functional is symmetric
with respect to the parity transformation $x\rightarrow -x$ and,
thus, the OP wave functions are either odd or even in the $x$
variable. The even solutions corresponding to the energy minimum
clearly satisfy the condition
\begin{eqnarray}
    \partial\Psi(\mathbf r)/\partial x|_{x=0}=0
    \label{bound_01}
    \end{eqnarray}
which coincides with the one imposed for the superconducting wedge
at the side with $\chi=\pi/2$. As a consequence, we find
$B_0^{*}(\pi/2)=H_{c3}^{\rm w}(\pi/2)\simeq 2H_{c2}$. The
variational parameters corresponding to this particular case
should be taken as follows:
$\alpha_1(\pi/2)=\alpha_2(\pi/2)=0.14,\,
\gamma_1(\pi/2)=\gamma_2(\pi/2)=-0.5,\, \eta(\pi/2)=0.6$.

However this mapping between the domain wall and the wedge, of
course, does not hold exactly for $\varphi_0\neq \pi/2$. Still
similarly to the wedge the shrinking of the wave function at small
angles $\varphi_0$ can result in the appearance of a vortex --
free solution (see Sec.~II~C).

{\it Two domain walls.} The above variational results for the
crossing domain wall and the sample edge remain valid also for two
domain walls crossing at rather large angle $\varphi_0\leq\pi/2$.
Due to the symmetry of the magnetic field profile, this
generalization is straightforward.

\section{Numerical simulations}
\begin{figure}[h]
    \begin{center}
    \includegraphics[width=7.cm]{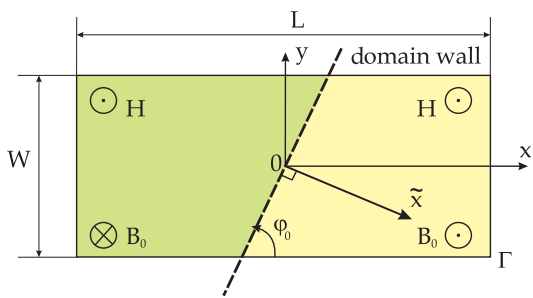}
    \end{center}
    \caption{(Color online) Schematic view of a F/S
    bilayer: a superconducting rectangle ($L\times W$) and
    a straight domain wall (dashed line) oriented at a certain angle $\varphi_0$
    with respect to the edges $y=\pm W/2$. $B_0$ is a
    stray field amplitude inside the domain,
    $H$ is an external magnetic field, the $\widetilde{x}$
    axis is chosen to be perpendicular to the domain wall, $\Gamma$ is the film boundary.}
    \label{System_numerical_analysis}
    \end{figure}

To confirm the findings obtained within the trial function
approach we proceed with the numerical analysis of the
superconductivity nucleation in a thin-film F/S bilayer within the
time -- dependent GL formalism. Let us consider a superconducting
thin-film rectangle in the plane $(xy)$ of the cartesian
coordinate system (having the lateral dimensions $L\times W$) in
the presence of a homogeneous external magnetic field $\mathbf H$
normal to the film plane and the step-like magnetic field
$\mathbf{b}(\widetilde{x})=\mathbf{z}_0B_0\,\rm
sign(\widetilde{x})$ of a straight domain wall oriented at a
certain angle $\varphi_0$ with respect to the film edges $y=\pm
W/2$ (see Fig.~\ref{System_numerical_analysis}).
    \begin{figure*}[htb!]
    \begin{center}
    \includegraphics[width=17.cm]{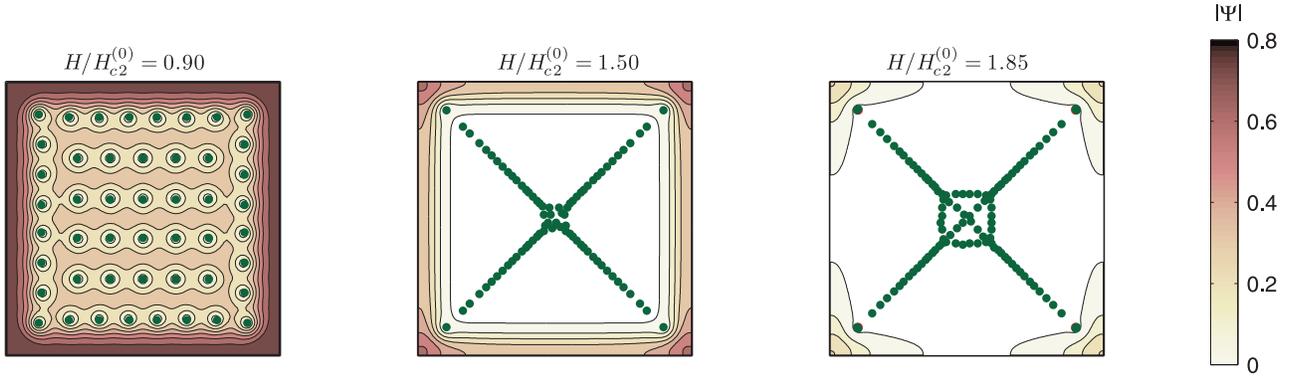}
    \end{center}
    \caption{(Color online) The OP distribution in a superconducting square
    ($20\xi_0\times 20\xi_0$) in an increasing external magnetic field
    $H$. The boundary condition (\ref{BoundCond_SC_I}) was imposed on the $\Psi$ function.
    Solid circles show the vortex chains.}
    \label{Fig_BoundCond_SC_I}
    \end{figure*}
    \begin{figure*}[htb!]
    \begin{center}
    \includegraphics[width=17.cm]{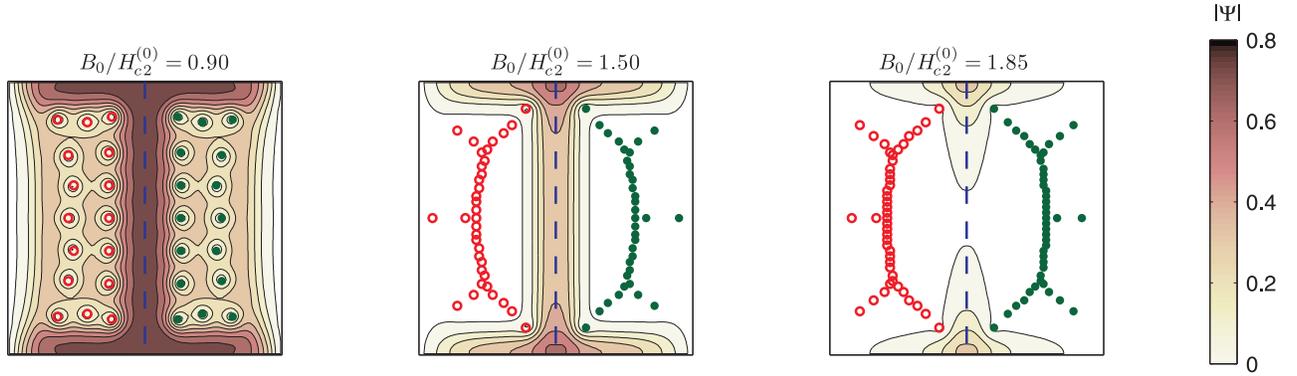}
    \end{center}
    \caption{(Color online) The OP distribution in a superconducting square
    ($20\xi_0\times 20\xi_0$) for an increasing amplitude $B_0$ of the magnetic stray field in the
    domains and $\varphi_0=\pi/2$. The boundary condition (\ref{BoundCond_SC_NM}) was used for $\Psi$ function in order to suppress
    the superconductivity nucleation near the corners of the sample. The vortex and antivortex chains are shown
    by solid and open circles, respectively.}
    \label{Fig-OP-Step}
    \end{figure*}

    \begin{figure*}[htb!]
    \includegraphics[width=17.cm]{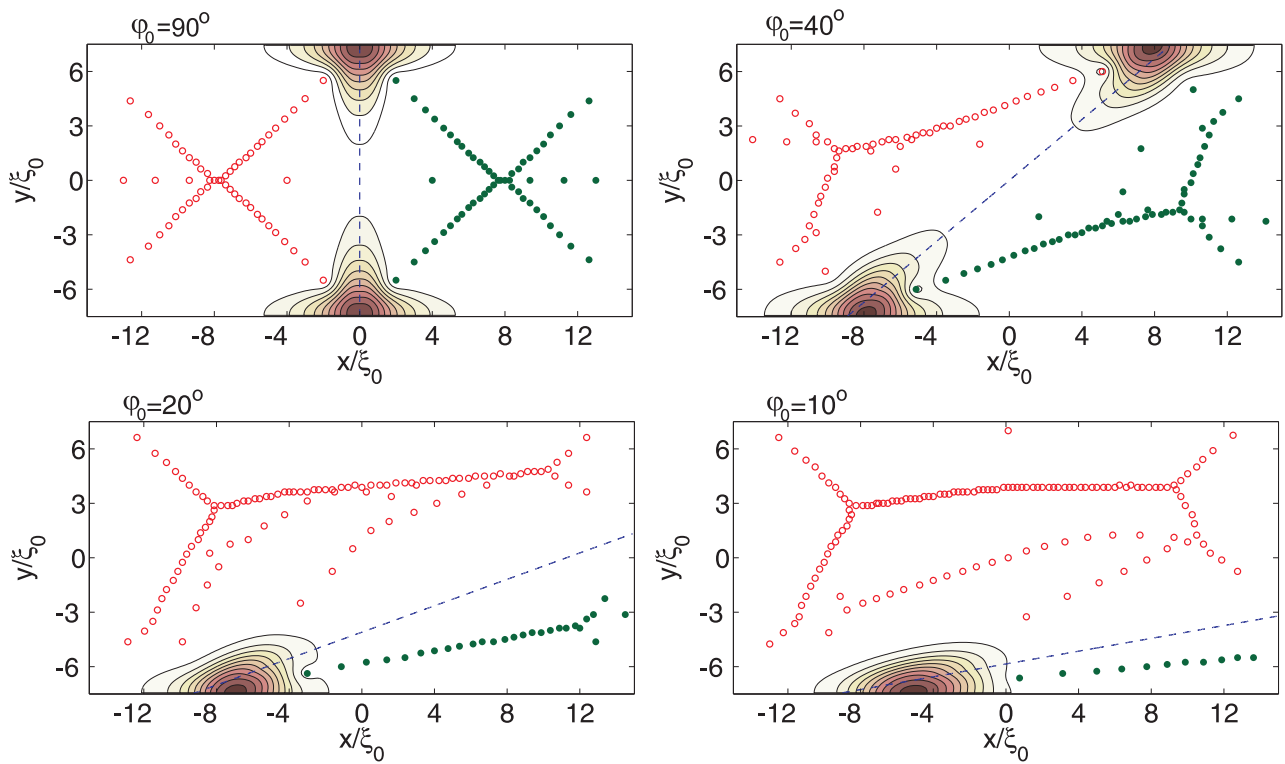}
    \caption{(Color online) Typical contour plots of the superconducting OP distributions for different $\varphi_0$ angles: $90^o$,
    $40^o$, $20^o$ and $10^o$. The vortex and antivortex chains are shown by the solid and open circles,
    respectively.} \label{OP_Angles}
    \end{figure*}

In order to obtain stationary OP distributions we simulate the
relaxation to an equilibrium state  on the basis of the time --
dependent GL model
    \begin{eqnarray}
    \nonumber -\eta \left[\hbar\frac{\partial}{\partial t}+2ie\,\Phi(\mathbf r)
    \right] \Psi(\mathbf r) = \alpha\,\Psi(\mathbf r) +
    \\
    \nonumber \frac{\beta}{2}\,|\Psi(\mathbf r)|^2 \Psi(\mathbf r) +
    \frac{1}{4m}\left[-i\hbar\nabla - \frac{2e}{c}\mathbf
    A(\mathbf r)\right]^2\Psi(\mathbf r) \ ,
    \\
    {\rm div}\, {\bf j}_n(\mathbf r) = - {\rm div}\, {\bf j}_s(\mathbf r) \ ,
    \label{Ginzburg -- Landau equations}
    \end{eqnarray}
where the parameter $\eta$ controls the rate of the OP relaxation,
$\Phi(\mathbf r)$ is the electrochemical potential, $\alpha$ and
$\beta$ are the parameters of the GL expansion, ${\bf j}_n(\mathbf
r)=- \sigma \nabla\Phi(\mathbf r)$ and
    \begin{eqnarray}
    {\bf j}_s(\mathbf r) = \frac{e}{2m}\left\{\Psi(\mathbf r)^*\left[-i\hbar\nabla -
    \frac{2e}{c}{\bf A(\mathbf r)}\right]\Psi(\mathbf r) +
    \mbox{c.c.}\right\}
    \label{Eq-2}
    \end{eqnarray}
are the densities of normal and superconducting currents, $\sigma$
is a normal conductivity, $\rm c.c.$ stands for complex
conjugation.  Focusing on the study of superconductivity
nucleation one can neglect the effect of vanishing supercurrents
on the magnetic field distribution and consider a fixed vector
potential profile corresponding to the magnetic field $B_z({\bf
r})=H + b_z(\widetilde{x})$.

The absence of normal current through the sample boundary imposes
a boundary condition on the potential $\Phi(\mathbf r)$:
    \begin{eqnarray}
    \nonumber \left.\frac{\partial\Phi(\mathbf r)}{\partial\bf n}\right|_{\Gamma}=0 \ ,
    \end{eqnarray}
where $\bf n$ is a unit vector normal to the edge $\Gamma$ of the
sample. The calculations have been made for two types of boundary
conditions on the superconducting OP wave function: (i)
superconductor -- insulator boundary condition
    \begin{eqnarray}
    \left.\big[-i\hbar\frac{\partial}{\partial\mathbf n}-
    \frac{2e}{c} \mathbf A_n\big]\Psi(\mathbf r)\right|_{\Gamma} = 0 \ ,
    \label{BoundCond_SC_I}
    \end{eqnarray}
and (ii) superconductor -- normal metal boundary condition
    \begin{eqnarray}
    \Psi(\mathbf r)\mid_{\Gamma}=0 \ .
    \label{BoundCond_SC_NM}
    \end{eqnarray}

To study the equilibrium phase diagram we need all transient
processes to be finished which corresponds to zero electrochemical
potential $\Phi(\mathbf r)=0$ and  zero time derivatives
$\partial\Psi(\mathbf r,t)/\partial t = 0$. In our calculations we
stopped the simulation procedure when the maximum of the
electrochemical potential reaches the accuracy of the numerical
computations ($10^{-15}\times\hbar |e| \alpha_0/2m\sigma\beta$ in
the dimension units), where $\alpha_0=|\alpha(T=0)|$. The
calculations have been carried out for a grid size $0.2\xi_0\times
0.2\xi_0$ and $\eta=10\times m\sigma\beta/\hbar e^2$. The time
interval between two subsequent iterations was chosen as follows:
$0.01\times m\sigma\beta/e^2\alpha_0$.

Let us start with the simplest case of a superconducting square
($L=W=20\xi_0$) placed only in a homogeneous magnetic field $H$.
We consider here the boundary conditions for $\Psi$ function in
the form~(\ref{BoundCond_SC_I}). By varying the external field $H$
we study the evolution of the OP distribution $\Psi(x,y)$ in the
superconducting square (see Fig.~\ref{Fig_BoundCond_SC_I}). One
can see that the Cooper pair wave function remains nonzero in the
vicinity of the sample corners at the field above $H_{c3}$.
Obviously, the corresponding nuclei at the corners can be
considered separately only for a rather large $L$ values well
exceeding the nucleus size. In this case each nucleus describes
the OP distribution in a superconducting wedge with a corner angle
$\pi/2$. The corresponding nucleation field $H_{c3}^{\rm w}\simeq
2H_{c2}$ appears to be in a good agreement with the results of our
variational analysis in Sec.~II~D. Our numerical simulations also
give evidence for the appearance of the vortex chains introduced
in the above consideration. These chains outgo from all four wedge
vertices along the corresponding bisectrices (see
Fig.~\ref{Fig_BoundCond_SC_I}).

We continue with the numerical analysis of the superconductivity
nucleation in a superconducting rectangle $L\times W$ affected in
the field of a straight domain wall. In order to suppress the
superconductivity nucleation near the sample corners  we use mixed
boundary conditions for the $\Psi$ function: the conditions
(\ref{BoundCond_SC_I}) and (\ref{BoundCond_SC_NM}) are taken at
the edges $y=\pm W/2$ and $x=\pm L/2$, respectively. Shown in
Fig.~\ref{Fig-OP-Step} is the transformation of the OP
distribution for a square with $L=W=20\xi_0$ $\varphi_0=\pi/2$
caused by the increase in the amplitude $B_0$ inside the domain.
The numerical results are in a full agreement with our analytical
findings in Sec.~II~D: (i) the superconductivity is localized near
the crossing of the domain wall and the sample edges $y=\pm W/2$
and survives up to the critical value $B_{0}^{*}=H_{c}^{\rm
w}\simeq 2H_{c2}$, (ii) we observe the vortex (antivortex) chains
outgoing from the crossing points of the superconducting channels.

Applying an external magnetic field $H$ we find out the phase
diagram for localized superconducting states. For a particular
case $B_0=1.5H_{c2}^{(0)}$ the corresponding phase transition line
is shown in the Fig.~\ref{Phase Diagram} by the dashed line. We
also observe the transition lines corresponding to the bulk
superconductivity (dotted line in Fig.~\ref{Phase Diagram}) and to
the domain wall superconductivity (dot-and-dash line in
Fig.~\ref{Phase Diagram}). Typical contour plots of the OP
distributions for different parts of the phase diagram (see
Fig.~\ref{Phase Diagram}) illustrate the switching between the
bulk and localized superconductivity nucleation.

In the limit of zero external field $H$ we have also analyzed the
dependence of the critical field amplitude $B_0^{*}$ on the
$\varphi_0$ angle for the sample sizes $L=30\xi_0$ and
$W=15\xi_0$. Typical plot of the dependence $B_0^{*}(\varphi_0)$
is shown in Fig.~\ref{Angular_dependence} by a solid line (2).
Both for small and large corner angles the numerical dependence is
in a good agreement with the results of the variational analysis
carried out in Sec.~II~D. Typical contour plots of the Cooper pair
wave function presented in Fig.~\ref{OP_Angles} show the
transformation of the vortex patterns with the changing
$\varphi_0$ angle.

Let's focus on the vortex (antivortex) arrangements presented in
Figs.~\ref{Phase Diagram}, \ref{Fig_BoundCond_SC_I},
\ref{Fig-OP-Step}, \ref{OP_Angles}. Near the crossing points of
the superconducting channels the vortex patterns obtained from
our numerical simulations appear to be in a good agreement with
the variational predictions for the infinite superconducting
channels (see Sec.~II) where the vortices (antivortices) form the
vortex (antivortex) chains. Far from these crossing points the
vortex distributions are more complicated and strongly depend on
the sample geometry and corresponding boundary conditions for the
Cooper pair wave function.

\section{Conclusion}
To sum up, we have investigated the distinctive features of
superconducting OP nucleation for the interacting superconducting
channels in strong magnetic field $H>H_{c2}$. We have studied
three generic problems: (i) the OP nucleation between two
superconductor/vacuum boundaries forming  a superconducting wedge;
(ii) the OP  nucleation between domain wall and the sample edge;
(iii) the OP nucleation between two domain walls. We have shown
that in all these cases the crossing of localized modes results in
the increase in the superconducting critical temperature. Using
both numerical and variational analysis of these problems we have
developed a description of the interference phenomena which govern
the structure of the OP patterns. The resulting critical
temperature enhancement and its magnetic field dependence should
be observable in resistive measurements of hybrid F/S structures.

\section{Acknowledgments}
This work was supported by the Russian Fund for Basic Research,
RAS under the Program ``Quantum physics of condensed matter",
Russian Agency of Education under the Federal Target Program
``Scientific and educational personnel of innovative Russia in
2009--2013", and by the Dynasty Foundation.

\end{document}